\documentclass[prb,showpacs,superscriptaddress,twocolumn,amssymb,amsfonts]{revtex4}
\usepackage{amsmath}
\usepackage{bm}
\usepackage{epsfig}
\usepackage{graphics}
\usepackage{dsfont}

\def\beq{\begin{equation}}
\def\eeq{\end{equation}}
\def\beqn{\begin{eqnarray}}
\def\eeqn{\end{eqnarray}}

\renewcommand{\bf}{\mathbf}

\begin{document}
\title{Quantum phase transition and fractional excitations in a topological insulator thin film with Zeeman and excitonic masses}
\author{Gil Young Cho}
\affiliation{Department of Physics, University of California,
Berkeley, CA 94720}
\author{Joel E. Moore}
\affiliation{Department of Physics, University of California,
Berkeley, CA 94720} \affiliation{Materials Sciences Division,
Lawrence Berkeley National Laboratory, Berkeley, CA 94720}
\date{\today}


%

\begin{abstract}
We study the zero-temperature phase diagram and fractional excitations when a thin film of 3D topological insulator has two competing masses: $T$- symmetric exciton condensation and $T$- breaking Zeeman effect. Two topologically distinct phases are identified: in one, the quasiparticles can be viewed as in a quantum spin Hall phase, and in the other a quantum anomalous Hall phase. The vortices of the exciton order parameter can carry fractional charge and statistics of electrons in both phases. When the system undergoes the quantum phase transition between these two phases, the charges, statistics and the number of fermionic zero mode of the excitonic vortices are also changed. We derive the effective field theory for vortices and external gauge field and present an explicit wave function for the fermionic zero mode localized at the excitonic vortices with or without orbital magnetic field.  The quantum phase transition can be measured  by optical Faraday or Kerr effect experiments, and in closing we discuss the conditions required to create the excitonic condensate.
\end{abstract}


\maketitle

\section{Introduction}
The surface states of three-dimensional topological insulators (TI) are two-dimensional metals with an odd number of Dirac points enclosed by the Fermi surface~\cite{fkm,mb,roy,hm}.  They avoid the so-called fermionic doubling problem and are known in several cases~\cite{fu1,qi1,JoelExciton} to support topological correlated states when different types of interactions are added.  The example motivating this paper is the topological exciton condensate formed by the Coulomb interaction between top and bottom surfaces of a thin film of topological insulator~\cite{JoelExciton}.  The exciton condensate on the TI surfaces carries electronic charge $\pm\frac{1}{2}$e at its vortices, i.e., half of the charge of the underlying electrons. This thin TI exciton condensate should be compared to that in a graphene bilayer, which has  similar relativistic electronic structure to a TI surface. The fractional properties are washed out by the spin/pseudospin degeneracies in graphene. For example, the Kekul\'e vortex on graphene can support  fractional quantum numbers~\cite{Kekule1, RyuVortex, MagneticKekule}, but the fractional excitation should be multiplied by $4$ (or $2$, depending on how the degeneracies are lifted) and results in integral quantum numbers as for electrons. (See also the charge fractionalization on the kagome lattice~\cite{GF} at filling $1/3$)

In the thin film, we have at least two Dirac fermions overall, one from the top surface and one from the bottom, and hence the system looks ``ordinary'' in the sense that we have an even number of fermionic degrees of freedom. However, it still has only half as many degrees of freedom as graphene, and it is known from previous work that the Topological Exciton Condensate(TEC) is distinct from an ordinary BEC of excitons.  It is fractional in the sense that it supports charge $\pm\frac{1}{2}$e at its vortices~\cite{JoelExciton}. This fractional quantum number of TEC results from the topological structure of the TI.  Now the TEC is $T$- symmetric but generates a mass at the Dirac points.  One way to motivate the present work is to compare this effect to the easiest way to open up a gap at the Dirac points of TI surface, namely, to break $T$-symmetry on the surface. There are a few natural ways to break $T$- symmetry and open up gaps at Dirac points: coating magnetic materials, or imposing magnetic fields etc. Thus if we introduce $T$- symmetry breaking perturbation and condense $T$- symmetric excitons on TI surfaces at the same time, then we expect that there should be an interesting interplay between TEC mass and $T$- symmetry breaking masses.  We find a quantum phase transition (QPT) between two phases that can be labeled by the behavior of their quasiparticles: Quantum Anomalous Hall (QAH) and Quantum Spin Hall (QSH) phases.  

\begin{figure}
\includegraphics[width=1\columnwidth]{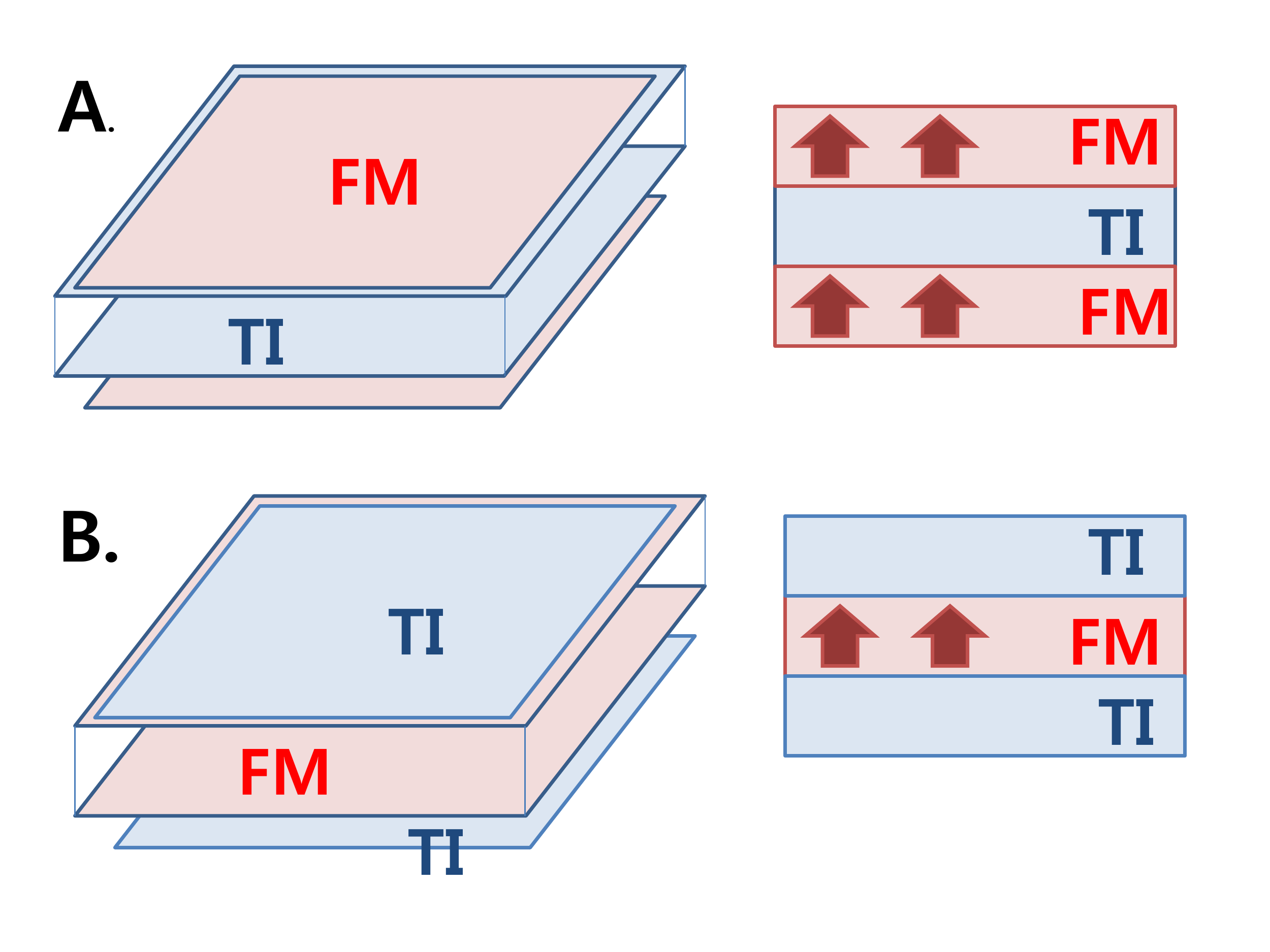}
\caption{A. Illustration of a thin topological insulator film sandwiched between ferromagnetic layers. To see the quantum phase transition between quantum spin Hall (QSH) and quantum anomalous Hall (QAH) phases, we need to align the magnetism along the same direction on both layers.  B. Alternate experimental scheme. We sandwich a thin ferromagnetic layer between two bulk topological insulators. If we put an insulator with small dielectric constant $\epsilon$ instead of ferromagnets, we expect to observe an increased critical temperature for the superfluid - insulator transition into the excitonic condensate.}
\label{Fig1}
\end{figure}

Here, we will first consider a thin TI film coated by ferromagnets (FM) on both sides (Fig.\ref{Fig1}). Then the system is gapped due to the $T$-breaking perturbation of the originally gapless surface. The electrons on the TI surface interact with the FM magnetization via Zeeman coupling $J_{H}\sigma \cdot M$. The $z$-component magnetization has a special role: it opens up a gap at the Dirac point with mass $J_{H}|M_{z}|$ sgn$(M_{z})$. This coupling is different from the case of graphene, where the pseudospins do not interact with magnetization and the Zeeman effect does not open up a gap. Other components of $M$ act as external gauge fields coupled to the Dirac fermions. When the system is fully gapped, we can safely integrate out fermions and obtain effective field theories for the gauge fields. We will develop the field theories for the fluctuating gauges and vortices in the two distinct phases. Even though the physics is most transparent when only the Zeeman coupling is considered, the same physics applies even when we consider an external $z$-directional magnetic field. The key difference in this case is that we have Landau levels(LL) instead of a Dirac spectrum. Despite this, we still have fractional vortices and a quantum phase transition when the Zeeman energy is comparable to TEC mass.

The following section studies the electronic band structure and shows that there is a QPT when the Zeeman mass is comparable to the exciton mass. At the QPT, the gap vanishes and the topological invariant (Chern number) of the bands is changed, which is why we refer to the transition as between QSH and QAH phases.  In Section III we obtain an effective field theory that captures the topological properties. From this we can derive information about the change  across the transition in the charge, statistics and the number of bound zero modes in a vortex. We next proceed to the uniform magnetic field case and find generally similar physics as in the Zeeman case. We also present the solution of the fermion zero mode at the vortices. In QSH, there is only one fermionic zero mode which disappears at the quantum phase transition to QAH. However, it's difficult to obtain the exact solution for the fermionic zero mode in the uniform orbital magnetic field but we show that we can manage to obtain analytic solutions for some limiting cases. In closing we discuss experimental detection of the phase transition and conditions required to observe the exciton condensate.

\section{Microscopic model and band structure}
We start from a microscopic model of a thin TI with a $T$-breaking Zeeman mass induced by proximity to a ferromagnet. We have two Dirac cones indexed by $\alpha=1,2$ (layer index) and short-ranged Coulomb interactions $U, V$. Here we specialize to the case $M_{x}\sim M_{y} \ll M_{z}$.  We include external EM probe fields $A_{i}$ which couple to the Dirac fermions and ignore the tunneling between layers, as this decays exponentially in thickness whereas the Coulomb energy decays only algebraically.  For simplicity we set the chemical potential $\mu =0$ (however, the topological properties studied here should be valid whenever the chemical potential is in the gap) and start from the weak Coulomb interaction where no phase transition is expected:
\begin{equation}
H = \sum_{\alpha=1,2}((-1)^{\alpha}v_{F} \bf{\sigma} \cdot (\bf{p} -e\bf{A})+ J_{H}M^{z}\sigma^{z} + U n_{\alpha}n_{\alpha}) + Vn_{1}n_{2} 
\label{1}
\end{equation} 
Here $\sigma$ are the Pauli matrices for (physical) spin.  Ignoring $U=\frac{e^{2}}{\epsilon l}$ (where $l$ is the system-dependent length of short-ranged Coulomb interaction) and $V=\frac{e^{2}}{\epsilon d}$(where $d$ is the width of the thin film), we can easily identify the eigenstates and eigenvalues of $H$. This is reasonable as $U$ and $V$ will only renormalize the Dirac velocity when they are small. From here onwards, we rescale $v_{F}= 1$ as this should not change topological properties. With this in mind, we rewrite the Hamiltonian in a more familiar format $H = F^{\dagger}{\bf H} F$ where $F$ is the four-spinor for the system:
\begin{equation}
{\bf H} = 
\left[ 
\begin{array}{cccc} 
m & p_{+} & 0 & 0 \\
p_{-} & -m & 0 & 0 \\ 
0 & 0 & m & -p_{+} \\
0 & 0 & -p_{-} & -m
\end{array} 
\right]
\label{2}
\end{equation}
There are two important points to note about this Hamiltonian ${\bf H}$. First, the Hamiltonian has the spectrum with the gap 2$|J_{H}M_{z}=m|$  at the Dirac point.
\begin{equation}
E_{p,s} = sgn(s)\sqrt{p^{2} + m^{2}}, \quad s = \pm 1
\end{equation}  
The spectrum is doubly degenerate due to the layer index. Moreover, from the matrix form(\ref{2}) of ${\bf H}$, we identify both up and down layers as in a QAH phase. To see this, let us take only the upper half of the block Hamiltonian ${\bf H}_{u}$ such that
\beq
{\bf H} = \left[ 
\begin{array}{cc} 
{\bf H}_{u} & 0 \\
0 & {\bf H}_{d}  
\end{array} 
\right], \quad
{\bf H}_{u} = 
\left[ 
\begin{array}{cc} 
m & p_{+} \\
p_{-} & -m  
\end{array} 
\right] = {\bf p}\cdot \sigma + m \sigma_{z}
\eeq
Now, we compute the winding number for ${\bf H}_{u} = {\bf p}\cdot \sigma + m \sigma_{z}$. We identify the vector ${\bf V}_{u}=(p_{x},p_{y},m)$ from ${\bf H}_{u}$, and compute the topologically invariant winding number (the first Chern Number) for ${\bf V}$ by evaluating 
\beq
C_{1} = \frac{1}{4\pi} \int dp_{x} \int dp_{y} {\hat V} \cdot \frac{\partial {\hat V}}{\partial p_{x}} \times \frac{\partial {\hat V}}{\partial p_{y}}
\eeq
where ${\hat V} = {\bf V}/|{\bf V}|$. The winding numbers for ${\bf H}_{u}$ and ${\bf H}_{d}$ are sgn($m$)$\frac{1}{2}$ which reflects the origin of the anomalous quantum Hall effect for the Dirac fermions. Taken as a whole, the system is in a QAH regime where the Hall conductance is quantized as $\frac{e^{2}}{h}$.

\begin{figure}
\includegraphics[width=1\columnwidth]{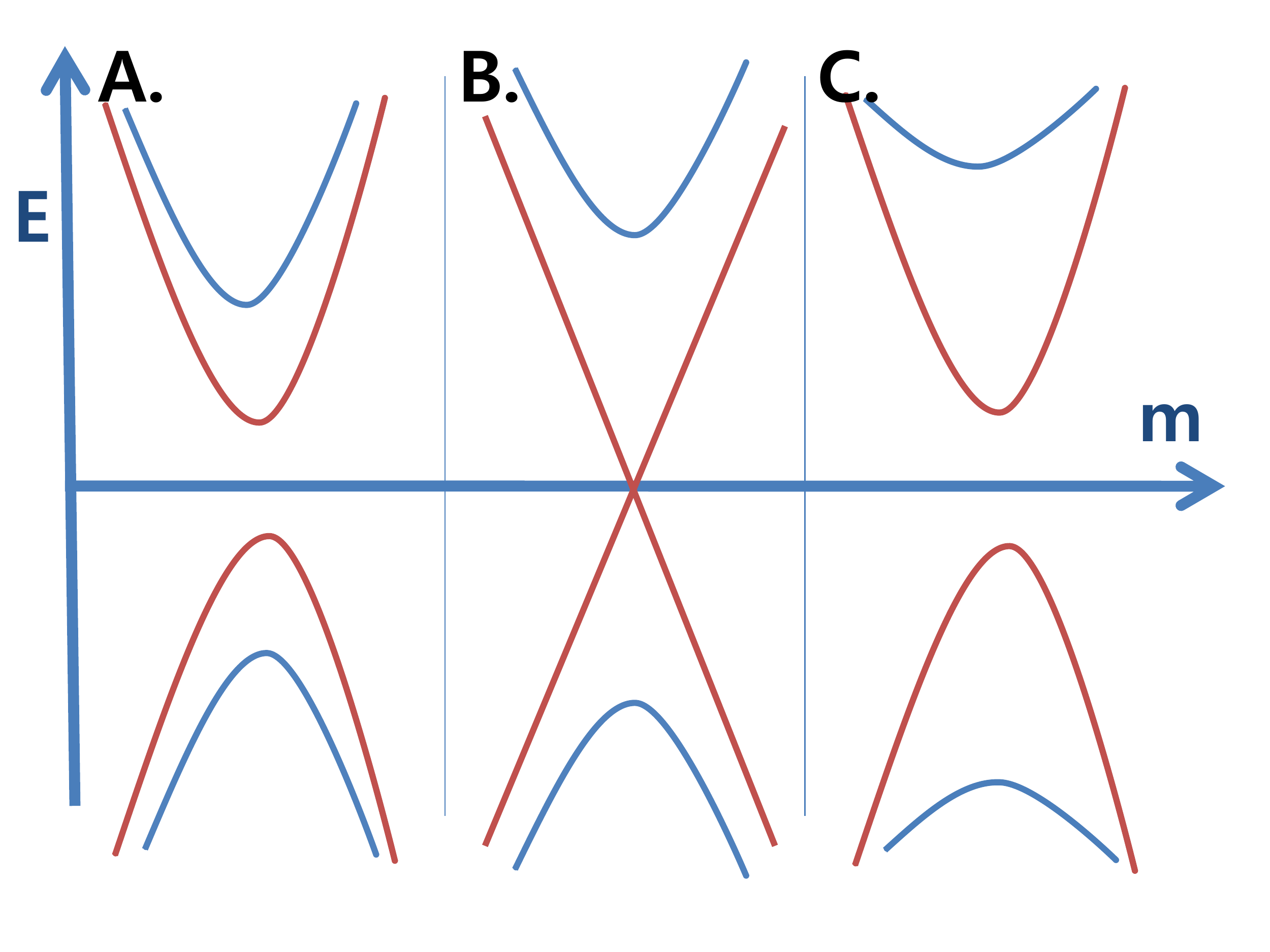}
\caption{Illustration of the dependence of band structures on the parameters $m>0$ and $\Delta>0$. Here, we assumed $\Delta \neq 0$. There are always $4$ bands, and the bands (blue) with the gap $m+\Delta$ never touches each other. The bands (red) with the gap $\Delta-m$ touch each other at $m=\Delta$, signaling the quantum phase transition. A. The band structure for the case $\Delta<m$. The system is gapped. B. When $\Delta=m$, the spectrum becomes gapless and there is a change in the winding number. C. For $\Delta >m$, the system is gapped again.}
\label{Fig2}
\end{figure}

So far, we have neglected the Coulomb interactions as they only renormalize parameters when they are small. However, when the Coulomb interaction increases, there can be phase transitions to a stripe phase~\cite{CDWTopologicalInsulator} for strong $U$, or exciton condensate~\cite{JoelExciton} for strong $V$. We are interested here in the TEC which is $T$- symmetric but still opens up a gap. To obtain TEC, we treat $V$ in terms of Mean Field Theory(MFT) in the exciton order parameter and  ignore $U$. We do the MFT for $Vn_{1}n_{2} =  f^{1\dagger}_{\sigma}M^{\sigma \rho} f^{2}_{\rho} + h.c$ with $M^{\sigma \rho} = <f^{1}_{\sigma}f^{2}_{\rho}{}^{\dagger}>$ (where $f^{1}{}_{\sigma}$ and $f^{2}{}_{\rho}$ are two-component Dirac fermions for layer $1$ of spin $z$-component $\sigma$ and $2$ of spin $z$-component $\rho$). Here, different $M$ describes different exciton condensates but we will choose an $M$ which opens up a gap at the Dirac points and is $T$- symmetric~\cite{JoelExciton}. The order is described by $M = \Delta \tau^{x}\sigma^{0}$ where $\tau$ are Pauli matrices acting on layer index, and $\sigma$ acts on spin as before. With the exciton order parameter, we have the following second-quantized mean-field Hamiltonian. 
\begin{align}
H &=  \sum_{\alpha}f_{\alpha}{}^{\dagger}((-1)^{\alpha} \sigma^{i} \cdot (p_{i} -eA_{i})+ m\sigma^{z}) f_{\alpha}\nonumber\\ 
&\qquad + \sum (f_{0}{}^{\dagger}\Delta \sigma^{0} f_{1} + h.c ) - \frac{\Delta^{2}}{V}
\end{align} 
In this form, the QPT between QAH and QSH phases is not obvious. To clarify the physics, we use the matrix forms of this Hamiltonian $H = F^{\dagger}{\bf H}F - \frac{\Delta^{2}}{V}$ as in the QAH case. 
\begin{equation}
{\bf H} = 
\left[ 
\begin{array}{cccc} 
m & p_{+} & \Delta & 0 \\
p_{-} & -m & 0 & \Delta \\ 
\Delta & 0 & m & -p_{+} \\
0 & \Delta & -p_{-} & -m
\end{array} 
\right]
\label{3}
\end{equation}
The first observation is that the spectrum of the Hamiltonian is always non-degenerate. Moreover, because $\sigma^{z}\tau^{0}$ commutes with $\sigma^{0}\tau^{x}$, we have masses $m-\Delta$ and $m+\Delta$ for the Dirac fermions. Explicitly, the spectrum is 
\begin{equation}
E(p,r,s) = sgn(s) \sqrt{p^{2} + (m-sgn(r)\Delta)^{2}},\quad s, r=\pm 1
\end{equation} 
Note that one of the masses changes its sign at $m=\Delta$, thus there is a QPT. Except at the QPT point, we have a fully gapped spectrum. Upon changing the sign of the mass, the system could change its chirality. To see that it does, we do a unitary transform to get a block-diagonal form of ($\ref{3}$):
\begin{equation}
{\bf H} = 
\left[ 
\begin{array}{cccc} 
(m+\Delta) & p_{+} & 0 & 0 \\
p_{-} & -(m+\Delta) & 0 & 0 \\ 
0 & 0 & (m-\Delta) & -p_{+} \\
0 & 0 &-p_{-} & -(m-\Delta)
\end{array} 
\right]
\end{equation}
From this representation, it is clear that we have a QPT at $m=\Delta$ as there is an abrupt change in the winding numbers at $m=\Delta$. Each block has the winding number sgn($m+\Delta$)$\frac{1}{2}$ and sgn($m-\Delta$)$\frac{1}{2}$. For $|\Delta|>|m|$, we have two cones of different chirality and thus we are in QSH. However, if we pass through $|\Delta|<|m|$, we have two cones of the same chirality and thus we are in QAH (See Fig.\ref{Fig2}). 

Note that our QSH phase does not support the helical edge state at the boundary of the sample (however, see the appendix) though the quasiparticles in the bulk low-energy theory can be considered in the standard QSH or two-dimensional topological band insulator phase. When the higher order kinetic energy correction to the Dirac Hamiltonian is included (e.g the hexagonal warping term $\sim \lambda (k_{+}^{3}+k_{-}^{3}) \sigma^{z}$), the winding numbers for QSH phase are changed from $\pm 1/2$ to zero for the both Dirac sectors. However, the low energy theory of our QSH phase is the same as that of the standard QSH physics.

In this section, we have seen from the band structure of the Hamiltonian \eqref{1} that there is a QPT between QSH and QAH phases. The QPT is driven by the competition of magnetic mass $m$ (which prefers the QAH phase) and exciton mass $\Delta$ (which prefers the QSH phase). At the transition, the mass gap vanishes and there is a change of the band structure's chirality. Except at the transition point, we always have a gap for the single-particle spectrum.  The goal of the next section is to obtain the key universal properties of the two phases in an abstract approach that is less dependent on microscopic details.

\section{Effective theory for vortices and fluctuations}
\begin{figure}
\includegraphics[width=1\columnwidth]{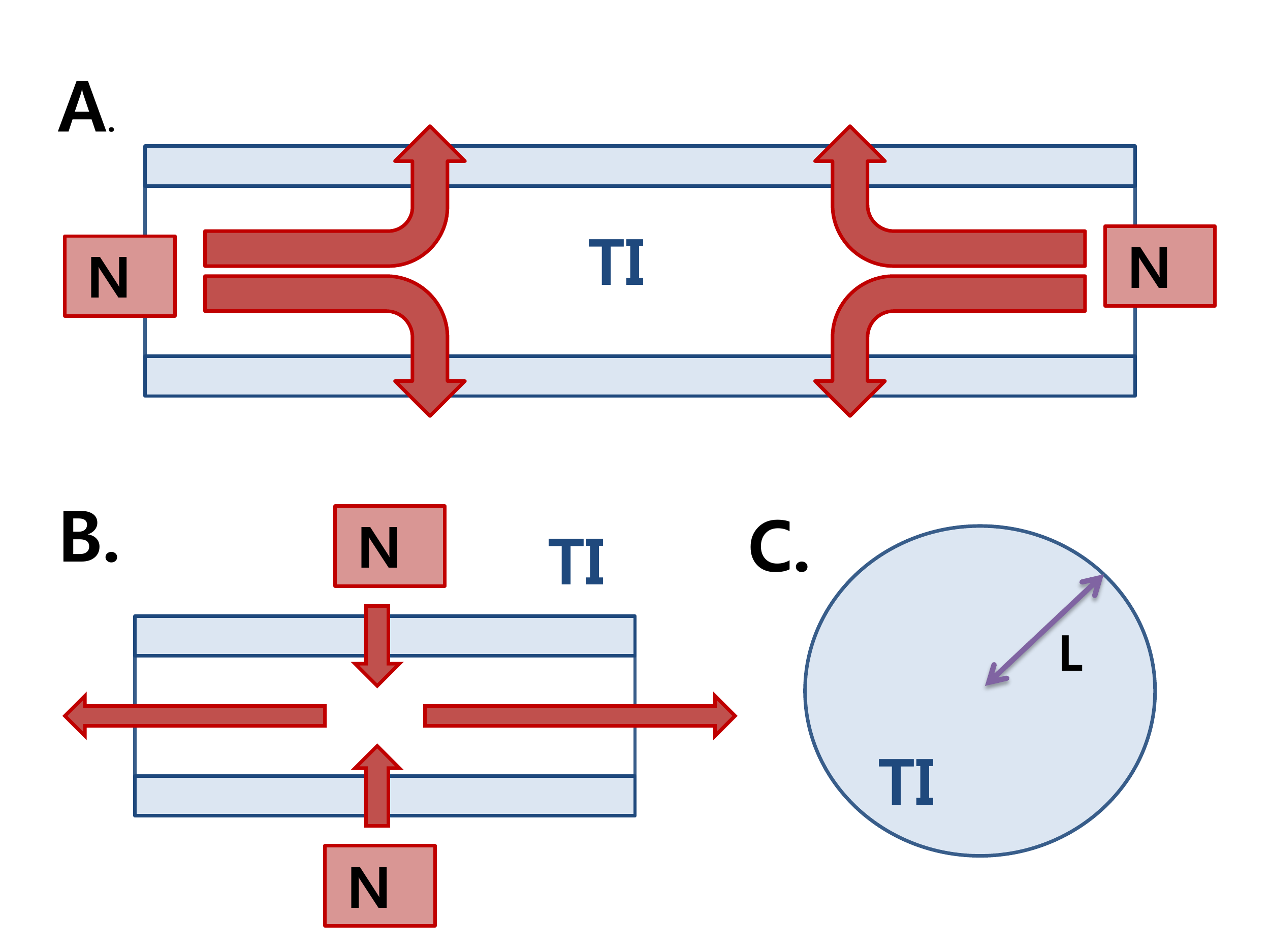}
\caption{Illustration of the generating axial gauge fluxes. General idea is to make magnetic field configuration which has $\int (B^{1}_{z}- B^{2}_{z})$ nonzero. A. Experimental scheme similar to that of the paper~\cite{FranzVortex}. We place magnets or solenoids at the boundary of the sample. B. Experimental scheme similar to that of the paper~\cite{Sonin}. We place solenoids on the top and below of the condensate. Without the solenoids, the vortex should create the magnetic flux itself to be screened. To estimate the energy of the vortex roughly (see the text), we consider that the flux enters at the vortex and leaves at the boundary of the sample. C. Top view of the scheme B. The exciton condensate has the radius $L$, and the solenoids are placed at the center of the condensate. }
\label{Fig4}
\end{figure}

Now that we have identified the chirality and chern numbers, we  integrate out the fermions to obtain  effective descriptions for the fluctuations and vortices of the order parameters and gauge fields. First, we identify which fields and vortices might appear in the effective theory. To do this, we include all possible interactions and write them as gauge fields acting on the Dirac fermions. We also include $L_{kin}(\Delta)$, is the kinetic term for $\Delta$.  
\begin{align}
H &= \sum_{\alpha}f_{\alpha}{}^{\dagger}((-1)^{\alpha} \sigma^{i} \cdot (p_{i} -eA^{\alpha}_{i})+ m\sigma^{z}) f_{\alpha}\nonumber\\
&\qquad+ \sum (f_{0}{}^{\dagger}\Delta \sigma^{0} f_{1} + h.c ) + L_{kin} (\Delta)
\end{align}
Here, $A^{\alpha}_{i}$ is the electromagnetic gauge field of the layer $\alpha = 1, 2$. From the gauge fields $A^{1}$ and $A^{2}$, we can define $A = A^{1}+A^{2}$ which couples to the total electromagnetic charge and $\beta = (A^{1}-A^{2})/2$ which couples to axial charge. Before getting into the details of the effective field theory, we study the nature of the vortex in the condensate. Especially, whether the vortex is accompanied with the gauge flux or not is crucial for the properties of the vortex. So, we begin with the kinetic term $L_{kin} (\Delta)$ for the exciton condensate. Due to the gauge symmetry of the exciton order parameter $\Delta = <f^{1}_{\sigma}{}^{\dagger}f^{2}_{\sigma}>$, the axial gauge $\beta_{\mu}$ couples to $\Delta$ minimally
\beq
L_{kin}(\Delta e^{i\phi}) \approx \frac{|\Delta|^{2}}{U}|\partial_{\mu} \phi - 2\beta_{\mu}|^{2} = \frac{\rho_{s}}{2} |\partial_{\mu} \phi - 2\beta_{\mu}|^{2}
\eeq
which is $XY$ - model (of the phase stiffness $\rho_{s}= |\Delta|^{2}/U$) coupled to the gauge field $\beta_{\mu}$.  Now, we imagine the vortex configuration of $\beta_{\mu}$ which can be generated by the solenoids placed near to the exciton condensate (Fig.\ref{Fig4}). The similar experimental scheme has been considered in the papers~\cite{FranzVortex, Sonin}. Then, the standard dual transformation allows us to write $\partial_{\mu}\phi - 2\beta_{\mu} \rightarrow 0$ for the vortex configuration of the exciton condensate. Hence, $\oint \beta = \pi {\bf Z}$ for the vortex in the condensate. This duality can be readily seen if we do the particle-hole transformation of the exciton order parameter $\Delta = <f^{1}_{\sigma}{}^{\dagger}f^{2}_{\sigma}>$ for the layer $2$ only. Upon the particle-hole transformation of the layer $2$, we see the exciton order $\Delta$ becomes the superconducting order parameter ($\Delta \rightarrow <f^{1}_{\sigma}{}^{\dagger}f^{2}_{\sigma}{}^{\dagger}>$). Thus, the exciton condensate is a `superconductor' in the axial channel and an insulator in the total charge channel as noted before~\cite{FranzVortex}. Here, if we have an axial vortex configuration for $\beta_{\mu}$, we have the counterflow (or an axial current) $J_{cf} \sim \rho_{s}\partial_{\mu} \phi$ encircling the vortex to screen the phase gradient due to $\beta_{\mu}$ (or vice versa). We estimate the screening length by 
\beq
L_{kin} = \frac{\rho_{s}}{2} |\partial_{\mu} \phi - 2\beta_{\mu}|^{2} + \frac{1}{2e^{2}}(\partial \beta)^{2} + \frac{1}{4e^{2}}(\partial A)^{2}
\label{Screening}
\eeq
where the last two terms are the kinetic energy for the gauge $\beta$ and $A$ which is obtained from the maxwell term for $A^{1}_{\mu}$ ($= \beta_{\mu}+ A_{\mu}/2$) and $A^{2}_{\nu}$ ($=\beta_{\nu} - A_{\nu}/2$). Hence, we identify an axial screening length $\lambda_{s} \sim 1/\sqrt{\rho_{s}e^{2}}$. Note that there's no screening length for total electromagnetic gauge field $A_{\mu}$ which reveals that we have an insulator in total charge channel. In principle, there should be another term $\sim (\beta_{\mu}/ed)^{2}$ in Eq.\eqref{Screening} when the film height $d$ gets thin enough. However, we ignore this term upon assuming that the film is not too thin $d > O(1)$. Though the energy of the vortex in the exciton superfluid could be reduced from infinity ($\sim \log L$ of the system size $L$) to finite value by screening, it might be costly for the gauge field $\beta_{\mu}$ for the vortex configuration. We will compare the energy cost for the gauge configuration of the screened vortex with the energy of the unscreened vortex and see when the vortex will tend to be screened by $\beta_{\mu}$. To do so, we roughly estimate the energy cost for the magnetic field configuration {\it spontaneously} driven (without the externally imposed magnetic fields in the condensate) by the screened vortex $\oint \beta = \pi$. We imagine that the magnetic field is dragged from infinity to the center of the circular exciton condensate of radius $L$ (with height $d$) and pulled out through the vortex positioned at the center of the condensate (See figures B. and C. of Fig.\ref{Fig4}). Magnetic field configuration ${\vec B}(r) = B(r) {\hat e}_{r}$ is 
\beq
B(r) = \frac{\Phi_{0}}{2\pi r \cdot d}
\eeq  
with the unit vorticity $\Phi_{0}$. The electromagnetic energy for this configuration $\int \frac{1}{2}\chi_{e} B^{2}$ grows as $\chi_{e} \Phi_{0}^{2}\log L$. Hence, if we compare the energy $\sim \rho_{s} \Phi_{0}^{2} \log L$ of the unscreened vortex and the energy $\sim \chi_{e} \Phi_{0}^{2}\log L$ of the screened vortex, we conclude to have screened vortices if the magnetic permiability of the exciton condensate $\chi_{e}$ is much smaller than the phase stiffness $\rho_{s}$. When this is not the case, there could be more interesting possibilities such as the {\it irrational} charge and statistics for vortices. We refer the reader for this discussion to the paper~\cite{RyuVortex} and the references there-in. On the other hand, if the magnetic field is supplied by the external mean such as the experimental set-up in our scheme (Fig.\ref{Fig4}) or Fig.1 of the papers~\cite{FranzVortex, Sonin}, it's much easier for the condensate to have the screened vortices. The vortex in the condensate can save {\it logarithmically divergent} energy cost ($\sim \log L$) by slightly tilting and trapping the magnetic fluxes which should take {\it finite} energy, and this concludes that the vortex in our experimental scheme will be always accompanied with the gauge flux $\oint \beta = \pi {\bf Z}$ (Fig.\ref{Fig4}). 

Another comment is in order. Even though the exciton order parameter $\Delta$ in the mean field theory is of the same form to the tunneling term between layers, tunneling term introduces a new term to the effective field theory $L_{tun} \sim -t \cos (\phi)$ as it tends to pin down the U$(1)$ phase of the exciton order parameter ($t$ is proportional to the tunneling strength between layers). This term breaks U$(1)$ symmetry and induces `sine-Gordon' theory coupled to the vortex matter which can be ignored as $t \rightarrow 0$ in our perspective~\cite{ZWang}.

Now that the allowed gauge couplings are identified, the fermions are integrated out to obtain the effective theory which is the main result of this section:
\begin{align}
L &= sgn(m+\Delta)\frac{\varepsilon^{\mu\nu\lambda}}{8\pi} (A + \beta)_{\mu}\partial_{\nu} (A + \beta)_{\lambda}\nonumber\\
&+sgn(m-\Delta)\frac{\varepsilon^{\mu\nu\lambda}}{8\pi} (A - \beta)_{\mu}\partial_{\nu} (A-\beta)_{\lambda} + O((\partial A)^{2})  
\label{5}
\end{align}
 (with $e=1$). Here, we ignore the Maxwell term as the Chern-Simons (CS) term dominates the low energy physics.  The appearance of the doubled CS theory is due to the two massive Dirac cones in Eq.\eqref{3} where we gauge one Dirac cone of mass $(m+\Delta)$ with $A_{\mu}+\beta_{\mu}$ and the other Dirac cone of mass $(m-\Delta)$ with $A_{\mu}-\beta_{\mu}$ and integrate out those gapped fermions to obtain the gauge theory of $\beta_{\mu}$ and $A_{\mu}$. This phenomenological approach reproduces the essential feature of Eq.\eqref{5} which is consistent with the previous study~\cite{RyuVortex}. We will also justify this effective theory Eq.\eqref{5} in the later part of this section. However, readers interested in a more detailed derivation of Eq.\eqref{EFT1} and Eq.\eqref{EFT2} from the Dirac hamiltonian are referred to the paper~\cite{RyuVortex} where this field theory is rigorously considered. In this field theory, we observe the coefficient of the CS term for $A$ changes at $m=\pm\Delta$. This signals the QPT at $m=\pm\Delta$. This simple theory captures nearly every essential property of the vortices and fluctuations of the system; the rich behavior of the vortices includes localized charges. 

For the QSH phase($|\Delta|>|m|$)), we have (up to the sign factor infront of the mutual CS theory for $\beta_{\mu}$ and $A_{\nu}$)
\beq
L = \frac{1}{2\pi} \varepsilon^{\mu\nu\lambda}\beta_{\mu} \partial_{\nu} A_{\lambda} + O(\partial A)^{2}
\label{EFT1}
\eeq
From this effective mutual CS (or equivalently $BF$) theory for $\beta_{\mu}$ and $A_{\nu}$, we can directly read off charge and statistics information for the exciton vortices. Here, it's interesting to notice that $BF$ theory emerges for QSH phase which reflects the time reversal symmetric nature of the phase. We now study the charge and statistics of the vortices. First, we see that the absence of the topological term for $\beta$ requires that the vortices have no statistics (statistical angle $\theta = 0$). However, the EM charge localized at the vortices is non-trivial in that  $q = \int \frac{\delta L}{\delta A} = \int \frac{1}{2\pi} \partial \beta =\frac{1}{2\pi}\oint \beta = \pm 1/2$, that is, we have a half of electron charge localized at the vortex, which is consistent with the previous consideration on the vortices of TEC ($\Delta=0$). In comparison with the paper~\cite{RyuVortex} and our result Eq.\eqref{EFT1}, we note that we miss the coefficient $sgn(\mu_{s})$ for $BF$ term which matters for the sign of the fractional charge at the vortex but this sign factor is not important for our thin film topological insulator problem. The electric charge of the fermionic zero mode at TEC is known to be defined {\it modular} integer~\cite{JoelExciton} in connection with the axionic $\theta$-vacuum i.e., the localized charge at the vortex is $-e(1/2 + n)$ for $n \in {\bf Z}$, and thus the sign of the fractional charge (or the sign of $BF$ term in Eq.\eqref{EFT1}) shouldn't be taken seriously. As was noticed in the earlier study of the fractional charge, the $\pm e/2$ charge is associated with the zero mode solution for the vortex. Therefore, we look for the solution of ${\bf H} \Psi = 0$ such that 
\begin{align}
{\bf H}&=\nonumber\\
&\left[ 
\begin{array}{cccc} 
m & (p_{+}-A_{+}) & \Delta e^{-i\theta} & 0 \\
(p_{-}-A_{-}) & -m & 0 & \Delta e^{-i\theta} \\ 
\Delta e^{i\theta} & 0 & m & -(p_{+}-A_{+}) \\
0 & \Delta e^{i\theta} & -(p_{-}-A_{-}) & -m
\end{array} 
\right]  
\label{ZeroModeSolution}
\end{align}
which is the Hamiltonian for the exciton vortex with vorticity $-1$. We solve this problem by the following ansatz $\Psi \sim (u_{1}, v_{1}, v_{2}, u_{2})^{T}$ with the constraint $u_{1}^{*} = - u_{2}$ and $v_{1}^{*}=v_{2}$~\cite{FranzVortex, JoelExciton}. Then, the single phase method provides $u_{1}\sim u_{2}^{*} \sim e^{-i\theta}$ and $v_{1}, v_{2}$ independent of the angular variable $\theta$~\cite{FranzVortex, JoelExciton}. With this in mind, we use the following ansatz~\cite{VortexSol1} 
\begin{align}
u_{1}(r,\theta) &= f(r) e^{i\pi/4} e^{-\int^{r} \Delta(r') dr'} e^{-i\theta},\nonumber\\
 v_{1}(r,\theta)&= g(r) e^{-i\pi/4}e^{-\int^{r}\Delta(r')dr'} 
\label{ansatz}
\end{align}
Upon substituting this ansatz, it is straightforward to solve the resulting differential equation and the solution is $u_{1} \sim \exp(-\int^{r} \Delta(r') dr') I_{0} (m r)$ and $v_{1} \sim \exp(-\int^{r} \Delta(r') dr')  I_{1}(m r)$ where $I_{n}$ is the $n$th modified Bessel function. This wavefunction is convergent if and only if we are in QSH phase $|m|<|\Delta|$. 

The statistical phase and localized charge of the vortex can be understood by the following adiabatic argument. In the QSH phase, we have two gapped Dirac fermions and they carry half-filled quantum Hall effect with the opposite chirality. When we thread the axial flux $\oint \beta = \pi$ adiabatically to the system which is identical to an exciton vortex of the circulation $2\pi$, each quantum Hall state collects charge $e/4$. Thus, the localized charge of the vortex is $e/2$ ($=e/4+e/4$). When we do the pair-wise exchange of the vortices, the statistical phase accumulated is $0$ ($=\pi/8-\pi/8$) because the statistical phase $\pi/8$ from one quantum hall state is precisely cancelled by the phase $-\pi/8$ from the other. Hence, the vortex in the QSH phase has no statistical angle: $\theta = 0$. 

In terms of the external EM gauge $A$, we have the response $\sim O(\partial A)^{2}$ which is the Maxwell kinetic term and non-topological. However, the topological property shows up in principle if we consider a particular type of the domain wall for the exciton mass $\Delta$.   (Appendix)

For the QAH phase($|\Delta|<|m|$), we have 
\beq
L = \frac{1}{4\pi} \varepsilon^{\mu\nu\lambda}A_{\mu} \partial_{\nu} A_{\lambda} +  \frac{1}{4\pi} \varepsilon^{\mu\nu\lambda}\beta_{\mu} \partial_{\nu} \beta_{\lambda} + O(\partial A)^{2}
\label{EFT2}
\eeq
From this effective theory, we now notice that we are in the quantized QAH from the leading CS term for the response of $A$. This tells us that we will have a circulating chiral edge state supporting the Hall conductance $e^{2}/h$ at the boundary of the sample.

On the other hand, we have no crossing term between $A$ and $\beta$ implying that the vortices of the exciton order will not carry any EM charge. On the other hand, the exciton vortices is `{\it anyonic}'. It looks like the coefficient of CS term implies that the vortices are fermionic, but this is not the case as we have $\oint \beta = \pm \pi {\bf Z}$ (In the usual CS theory $L = \frac{K}{4\pi} a \partial a$, $\oint a = 2\pi$ defines the unit vortex). With this consideration, we now have fractional statistics between vortices with the statistical angle $\theta = \pi/4$ (half of semionic statistics). Note that the vortex in QAH phase does not carry a zero mode solution for the equation $\eqref{ZeroModeSolution}$. However, we can still talk about the exciton vortex as it is quantized to have circulation $2\pi {\bf Z}$ and well-defined excitation. This anyonic vortex emergent in QAH phase can be considered as an example of anyons from `weakly' interacting systems and this is similar to the previous study on integer quantum hall effect with the fully filled lowest landau level adjacent to the type-$II$ superconducting film~\cite{MF}.

The behavior of the vortex can be understood in terms of the adiabatic argument as before. In the QAH phase, the two sectors of gapped Dirac fermions have the same chirality. Thus, the axial gauge flux $\oint \beta = \pi$ associated with an exciton vortex will collect $e/4$ in one sector and $-e/4$ in the other. This gives the total charge $0$ ($=e/4-e/4$). If we exchange a pair of vortices, we now have the statistical phase $\pi/4$ ($=\pi/8+\pi/8$) by adding up the phases accumulated from the two sectors. 

We now take a different point of view to present the justification of our theory Eq.\eqref{5}. We base on the adiabatic argument and physical understanding of the fractional charge and statistics of vortices. First, we expect that there could be (at most) two CS terms for the effective field theory of $\beta_{\mu}$ (axial gauge) and $A_{\mu}$ (electromagnetic gauge) due to the two massive Dirac cones
\beq
L_{eff} = \frac{C_{1}}{2\pi} \varepsilon^{\mu\nu\lambda} A_{\mu}\partial_{\nu}\beta_{\lambda} +  \frac{C_{2}}{4\pi} \varepsilon^{\mu\nu\lambda} A_{\mu}\partial_{\nu}A_{\lambda} + \frac{C_{3}}{4\pi} \varepsilon^{\mu\nu\lambda} \beta_{\mu}\partial_{\nu}\beta_{\lambda}
\label{e1}
\eeq 
and the effective field theory should behave well because of the gap except the critical point, i.e., the coefficients $(C_{1},C_{2},C_{3})$ in Eq.\eqref{e1} change only at the critical point $m = \Delta$. So our basic strategy is to extract the coefficients for QSH and QAH phases. As $T$-symmetry has the important role, we study the transformation of $\beta_{\mu}$ and $A_{\mu}$ under $T$-symmetry operation: $(\beta_{0}, \beta_{i}) \rightarrow (-\beta_{0}, \beta_{i})$ and $(A_{0}, A_{i}) \rightarrow (A_{0}, -A_{i})$. Thus, the first $BF$ term in Eq.\eqref{e1} between $\beta_{\mu}$ and $A_{\nu}$ is $T$- symmetric, and the other two CS terms are $T$-breaking.  

For QSH phase, the phase should respect $T$-symmetry in the effective field theory Eq.\eqref{e1} though we break $T$-symmetry microscopically by inclusion of magnetization mass $\sim m \sigma^{z}$. We can extract the coefficient $(C_{1},C_{2},C_{3})$ for Eq.\eqref{e1} at $m=0$ with the fixed $\Delta$ because the QSH phase with finite $|m|<|\Delta|$ is adiabatically connected to the system of $m=0$ without closing the gap. This consideration gives $C_{2}=C_{3}=0$ for the entire range of QSH phase. And we supplement the field theory with the information that the vortex $\oint \beta_{\mu}= \pi$ carries the fractional charge $\pm e/2$ with finite $\Delta$ at $m \rightarrow 0$~\cite{JoelExciton}. This translates to the field theory as $C_{1}=1$. Hence, we have the effective field theory for QSH phase $L_{eff} = \frac{1}{2\pi} \beta\partial A $ (with the implicit antisymmetrization of indices for fields) and this is precisely the same as the field theory we obtained before Eq.\eqref{EFT1}. 

On the other hand, QAH phase has the total Chern number which fixes $C_{2}=1$. For $C_{1}$ and $C_{3}$, we utilize the adiabatic argument with $\Delta \rightarrow 0$ while keeping $m$ finite. At this limit, the axial gauge flux $\oint \beta = \pi$ associated with an exciton vortex will collect $e/4$ in one layer and $-e/4$ in the other. This gives the total charge $0$ for the vortex which translates $C_{1}=0$ for the effective field theory Eq.\eqref{e1}. When we do a pair-wise exchange of vortices $\oint \beta = \pi$, we now have the statistical phase $\pi/4$ ($=\pi/8+\pi/8$) by adding up the phases accumulated from the two layers. This fixes $C_{3}=1$. As a whole, we obtain the effective field theory for QAH phase $L = \frac{1}{4\pi} A\partial A +  \frac{1}{4\pi}\beta\partial \beta$ (with the implicit antisymmetrization of indices for fields) which is the same as in Eq.\eqref{EFT2}. Finally, it's straightforward to rewrite these field theories in both phase into a sinlge theory of the form Eq.\eqref{5}.  

In the paper~\cite{FranzVortex}, it's shown that the excitonic vortex may carry the {\it irrational} statistics due to the irrational axial charge $\delta Q_{v} = Q^{1}-Q^{2}$ bound to the vortex. (Note that in the bilayer system, the fluctuation in the axial charge could be finite unlike the total charge which is suppressed by the Coulomb interaction between layers) With this additional axial charge and the axial $\pi$ flux of the vortex, the statistical angle is given as $\pi\delta Q_{v}$ which could be continuously tuned~\cite{FranzVortex}. In our case, this might not be the case. First of all, there's no term breaking the symmetry between layers (such as the voltage drops $\mu_{s}$), we don't have any irrational statistics as the axial charge is $\delta Q_{v} = 0$ due to the symmetry between up and down layers. We also can deduce the form of the effective field theory from the adiabatic argument when the voltage drop $\mu_{s}$ between layers is included in consideration. When the voltage drop $\mu_{s}$ is slowly turned on from $0$ in the system, we evolve the mass $\Delta$ in Eq.\eqref{EFT1} and Eq.\eqref{EFT2} into $\sqrt{\Delta^{2}+\mu_{s}^{2}}$ without closing gaps (the total masses for the two Dirac cones are $m \pm \sqrt{\Delta^{2}+\mu_{s}^{2}}$). So we predict at least in our effective field theory and the adiabatic argument that the statistics and charge of vortices are fractionalized and quantized as in the case without $\mu_{s}$ .  

In the QAH phase, the system acquires a topological QAH response to the external EM gauge. This has a direct implication on the Faraday angle $\theta_{F}$ which is related to Hall conductance $\sigma_{xy}$ of the system~\cite{TseMacDonald1,TseMacDonald2,qi1,QiZhang2}. In QAH phase, we expect to have a strong Faraday effect as $T$-symmetry is broken. On the other hand, we wouldn't have a strong Faraday effect in QSH phase as we do not break $T$-symmetry {\it effectively} though we break $T$-symmetry microscopically. This optical response could be used in experiment to distinguish the two phases. In a thin film with  interlayer coherence (exciton order), the electron tends to be ambiguous on its layer index. Thus, we can think of the thin film as like a {\it single} layer with two Dirac cones. Then, as pointed out from the previous section, we have two Dirac cones with masses $m+\Delta$ and $m-\Delta$, and the two Dirac cones contribute to total $\sigma_{xy}$ equally. We specialize for  low frequency $\omega<<E_{c}$ ($E_{c}$ is the cut-off for the energy, and typically we can take $E_{c}$ as the bulk gap~\cite{TseMacDonald1}) and the chemical potential lying in the gap. Then, at the leading order $O(\omega/E_{c}, m/E_{c}, \Delta/E_{c})$ 
\beq
\sigma_{xx}=0, \quad \sigma_{xy} = \sigma_{xy},{}_{+} + \sigma_{xy},{}_{-}
\eeq
where $\sigma_{xy},{}_{\pm} =$ sgn$(m\pm\Delta) \times \frac{\alpha}{4\pi} (1-\frac{|m\pm\Delta|}{E_{c}})$ (Here, $\alpha$ in the coefficient is the fine constant, i.e $\alpha = 1/137$). For convenience, we take the Zeeman mass $m$ and exciton order parameter $\Delta$ positive. In the QAH phase, we have $\sigma_{xy} = \alpha/2\pi $ with the limit $m/E_{c} \rightarrow 0$. This gives $\theta_{F} = \tan^{-1} (\frac{2\alpha}{\sqrt{\epsilon'/\mu'}+\sqrt{\epsilon/\mu}}) \sim 10^{-3}$ rad, which is simply the double of the previously studied on the single-layer Dirac cone with $T$- breaking. In the QSH phase, where we preserve $T$-symmetry effectively, we have totally different behavior in that $\sigma_{xy} = \frac{\alpha}{4\pi} \times \frac{2m}{E_{c}} \rightarrow 0$ in the limit $m/E_{c} \rightarrow 0$. Thus, there is no significant $\theta_{F}$ even though $T$-symmetry is broken at a microscopic level. The small $T$-breaking shows up only in the order of $O(m/E_{c})$; $\theta_{F} \sim \alpha \times O(m/E_{c})$. If we plug $m\sim$ 10 meV and $E_{c}\sim$ 0.3 eV for Bi$_{2}$Se$_{3}$, then we have $\theta_{F} \sim \alpha \times O(m/E_{c}) < 10^{-4}$ rad, much smaller than $\theta_{F}$ for QAH phase.

\section{Uniform orbital magnetic fields in a thin film}
\begin{figure}
\includegraphics[width=1\columnwidth]{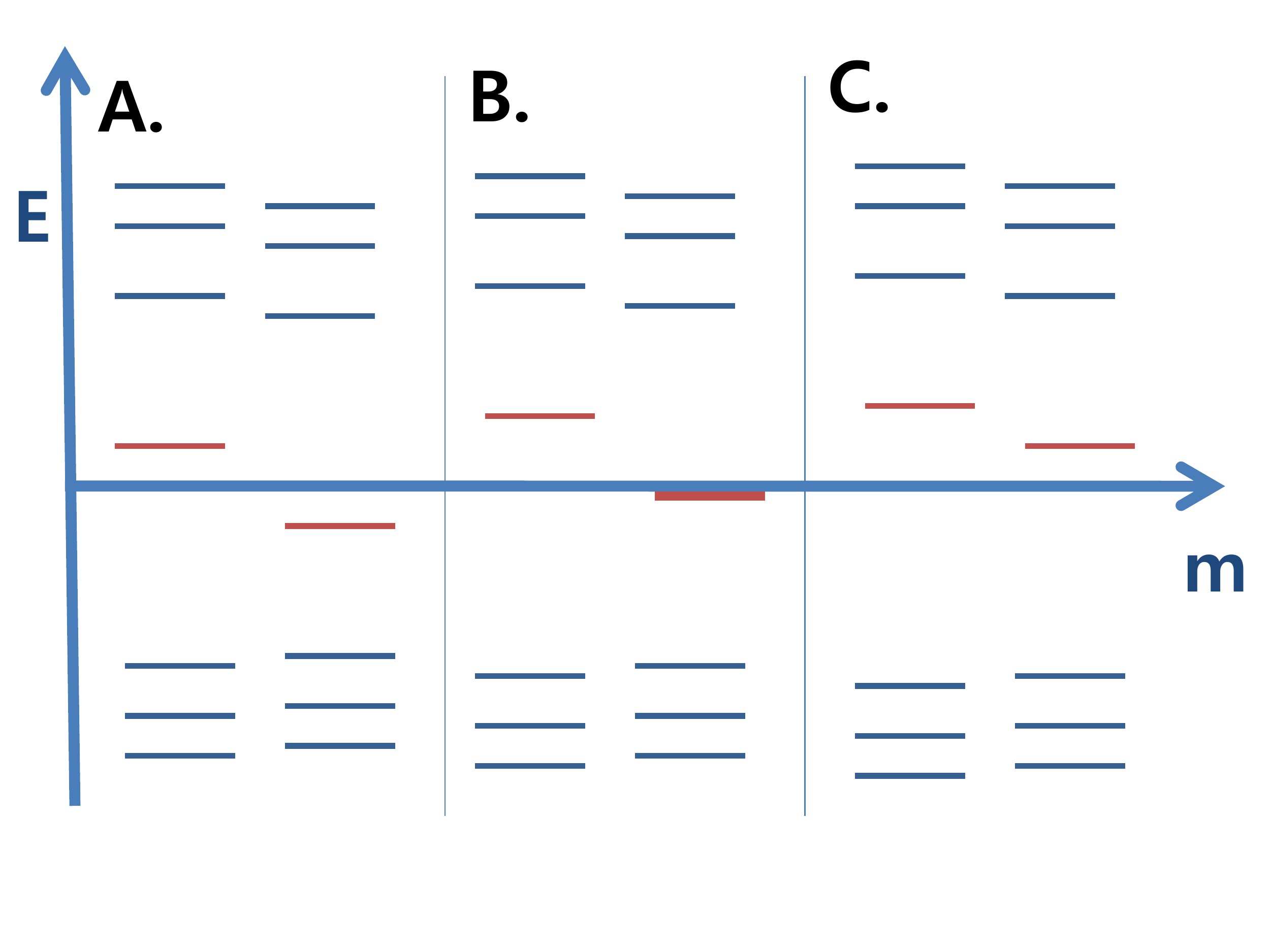}
\caption{Illustration of the Landau Levels depending on the parameter $m>0$ and $\Delta>0$. Here, we assumed $\Delta \neq 0$. We always keep the chemical potential $\mu$ at $E=0$, and the $0$th LL passes through the fermi energy $E=0$ at $m=\Delta$. The levels of LL index $n$ with $n \neq 0$ never change their filling as their energies are always positive (if $n>0$) or negative (if $n<0$). These levels (blue) do not play an important role. On the other hand, the two $0$th LLs drive the quantum phase transition at $m=\Delta$ because the sign of their energies are dependent on $m$ and $\Delta$. The filling of the lower $0$th LL (red) in the figure is abruptly changed at $m=\Delta$ as it passes through $\mu$. A. the LLs of the case $\Delta<m$. The lower $0$th LL is fully- filled B. when $\Delta=m$, the $0$th LL experiences change in its LL filling. C. Passing $\Delta >m$, the $0$th LL is fully emptied}
\label{Fig3}
\end{figure}

We have seen that the interplay between Zeeman and exciton masses induces interesting physics on the vortices and in the electromagnetic  response. However, there is another natural way to break the $T$- symmetry: uniform magnetic fields along $\hat{z}$. We have Zeeman interaction due to magnetic field $B$ via $\sim g \sigma_{z} B$ but we also have Landau Levels (LL). We will see that the zeroth LL will determine the physics and we can obtain the same effective theory as before. The appearance of the same effective theory can be traced back to the CS effective theory for QHE. Now, the coefficient of the effective theory is decided by the filling of the LLs, rather than the winding numbers of the band structures, and the phase transition at $m=\Delta$ is replaced by the quantum hall phase transition where the filling of LLs are suddenly changed. We begin with LLs of Dirac fermions which is similar to graphene. But the crucial difference here is the Zeeman coupling and degeneracy. Under the uniform magnetic field $B$, electrons form the LLs with index $N \in {\bf Z}$ 
\begin{align}
E(N,r) &= sgn(N) |v_{F}| \sqrt{C|N| +(m +sgn(r) \Delta)^{2}}\nonumber\\
 &\qquad\qquad\qquad\qquad\qquad\qquad r = \pm 1, N \neq 0
\end{align} 
and $C = \frac{ehB_{z}}{\pi c}$. The zeroth LL is sensitive to the competition between $m$ and $\Delta$ in that (See also Fig. \ref{Fig3})
\begin{equation}
E(N=0, r) = m + sgn(r)\Delta, \quad r =\pm 1
\end{equation}
We begin with the case where the chemical potential lies at $E_{F}= 0$ as before. Then, we see that every LL with negative (positive) index $N$ is always filled (empty) independent of the parameters $B, m$ and $\Delta$. However, the filling of the zeroth LL is dependent of the parameters (See Fig.\ref{Fig3}). For $m>\Delta$, we have that both of zeroth LLs are empty (as they are above the chemical potential $\mu=0$). Thus, we have QHE with the quantized hall conductance $\frac{e^{2}}{h}$ (Note that there is an offset by $1/2$ of the QHE for Dirac fermions). On the other hand, if we have $m<\Delta$, one LL is filled and the other is empty so we have a spin-Hall-like response. For the QHE, CS theory is the effective field theory 
\begin{equation}
L_{QHE}= \frac{1}{4\pi K} \varepsilon^{\mu\nu\lambda} A_{\mu} \partial_{\nu} A_{\lambda}  
\end{equation} 
where $K$ is the inverse of the filling (i.e $K = 1/\nu$). Hence, we conclude the effective theory for the system as $K=\pm 2$  (The sign of $K$ is determined by $m$ and $\Delta$)
\begin{align}
L_{QHE} &= sgn(m-\Delta) \frac{\varepsilon^{\mu\nu\lambda}}{8\pi} A_{1}{}_{\mu} \partial_{\nu} A_{1}{}_{\lambda} \nonumber\\
&\quad +sgn (m+\Delta)\frac{\varepsilon^{\mu\nu\lambda}}{8\pi} A_{2}{}_{\mu} \partial_{\nu} A_{2}{}_{\lambda} +O((\partial A)^{2}) 
\end{align} 
This is exactly the same effective field theory before \eqref{5}, and what all we need to do is to assign the correct coupling of axial gauge $\beta$ for each of the zeroth LLs.  The two LLs are from the two Dirac cones of the opposite axial charge $\pm 1$ and the same EM charge $+1$, i.e., $A_{1} = A + \beta$ and $A_{2}= A-\beta$. Upon plugging this in, we restore the same effective field theory Eq\eqref{5} as in the previous section. Thus, we have the same charges and statistics if we include the vortex fields in the effective theory. We see immediately that much of the physics studied in the previous section applies in this case. 


However, the zero energy solution for the magnetic field case is not trivial to generalize as the uniform magnetic field cannot be gauged away. Furthermore, this unifrom orbital magnetic field destroys the particle-hole symmetry of the Hamiltonian, and thus we cannot use the same ansatz Eq.\eqref{ansatz} with $\Psi \sim (u_{1}, v_{1}, v_{2}, u_{2})^{T}$ with the constraint $u_{1}^{*} = - u_{2}$ and $v_{1}^{*}=v_{2}$. This form of the ansatz~\cite{FranzVortex, VortexSol1} relies on the anticommutation relation $\{C, H\}=0$ where $C = i\tau^{y}\sigma^{y} K$ and $K$ is the complex conjugation. With the orbital magnetic field, we no longer have the anticommutation relation $\{C, H\}=0$. This $\{C, H\}\neq 0$ manifests in the LL spectrum (See Fig.\ref{Fig3}). So we instead make an `educated guess' for the ansatz $\Psi \sim (u_{+}, v_{+}, v_{-}, -u_{-})^{T}$ and 
\begin{align}
u_{\pm}(r,\theta) &= f_{\pm}(r) e^{\pm i\pi/4} e^{-\int^{r} \Delta(r') dr'} e^{-i\theta},\nonumber\\
 v_{\pm}(r,\theta)&= g_{\pm}(r) e^{\mp i\pi/4}e^{-\int^{r}\Delta(r')dr'} 
\label{ansatz2}
\end{align}
With this ansatz, we obtain four independent linear equations of $f_{\pm}(r)$ and $g_{\pm}(r)$
\begin{align}
m f_{+} &+ \Delta g_{-} + (\frac{\partial}{\partial r} + \frac{Br}{2}) g_{+} = 0 \nonumber\\
m g_{+} &+ \Delta f_{-} + (\frac{\partial}{\partial r} +\frac{1}{r}- \frac{Br}{2}) f_{+} = 0 \nonumber\\
m g_{-} &+ \Delta f_{+} + (\frac{\partial}{\partial r} +\frac{1}{r} + \frac{Br}{2}) f_{-} = 0 \nonumber\\
m f_{-} &+ \Delta g_{+} + (\frac{\partial}{\partial r} - \frac{Br}{2}) g_{-} = 0 
\label{SysOde}
\end{align}
We can try to solve this system of differential equations by reducing into the following two equations via canceling $f_{-}$ and $g_{-}$ in the above equations. 
\begin{align}
(M^{2} + \partial^{2} +\frac{2}{r}\partial - \frac{B}{2}-\frac{B^{2}r^{2}}{4})f_{+} + m (2\partial +\frac{1}{r}+ Br)g_{+}&=0 \nonumber\\
 (M^{2} + \partial^{2} + \frac{B}{2}-\frac{B^{2}r^{2}}{4})g_{+} + m (2\partial+\frac{1}{r} - Br)f_{+}&=0
\label{SysOde2}
\end{align}
where we abbreviate $\frac{\partial}{\partial r} = \partial$ and $M^{2} = m^{2}-\Delta^{2}$. The above differential equation Eq.\eqref{SysOde2} is difficult to solve and not clear if they admit analytic solutions without approximations. Rather than directly attempting to solve the system of differential equations, we look for a few solvable limits with the analytic solutions in terms of the confluent hypergeometric functions. The most convenient and important limit is when the zeeman coupling vanishes ($m \rightarrow 0$) which effectively decouples $f_{\pm}$ and $g_{\pm}$. In this limit, the solution for Eq.\eqref{SysOde2} is $f_{\pm}=0$ and 
\beq
g_{+} (r) \sim \frac{1}{\sqrt{r}} W (\frac{1}{4}- \frac{\Delta^{2}}{2B};-\frac{1}{4}; \frac{Br^{2}}{2})
\label{Sol1}
\eeq
where $W(\frac{1}{4}- \frac{\Delta^{2}}{2B};-\frac{1}{4}; \frac{Br^{2}}{2})$ is the Whittaker function~\cite{VortexHypergeometric}, or equivalently we can represent it as the parabolic cylinder function $g_{+}(r) \sim D_{p}(\frac{Br^{2}}{2})$ with $p = -\frac{\Delta^{2}}{2B}$ which appears in the Kekule vortex solution in the graphene~\cite{MagneticKekule} under the uniform orbital magnetic field. This limit of the solution corresponds to the case $|\Delta| >> |m|$, i.e., the negligible zeeman coupling. 

The other important limit where we can obtain an analytic solution for Eq. \eqref{SysOde} is when $r\rightarrow \infty$. In this limit, we ignore the potential terms of the order $\sim O(1/r)$  for $f_{\pm}$ and $g_{\pm}$ in Eq.\eqref{SysOde}. From this approximation, we try to see if there's a convergent solution for the system of differential equations. In the limit $r \rightarrow \infty$, we obtain 
\begin{align}
g_{+}(r) \sim \frac{1}{\sqrt{r}} \{&C_{1}W (\frac{1}{4}- \frac{(\Delta-m)^{2}}{2B};-\frac{1}{4}; \frac{Br^{2}}{2}) \nonumber\\ 
 &+ C_{2}W (\frac{1}{4}- \frac{(\Delta+m)^{2}}{2B};-\frac{1}{4}; \frac{Br^{2}}{2})\}
\label{Sol2}
\end{align}
with the initial condition dependent coefficients $C_{1}$ and $C_{2}$. It's now obvious that the solution $g_{+}(r)$ in Eq.\eqref{Sol2} reduces to $g_{+}(r)$ from Eq.\eqref{Sol1} when $m/|\Delta| \rightarrow 0$. As $g_{+}(r) \sim r^{2M} \exp(-Br^{2}/4) (1+ O(1/r^{2}))$ as $r\rightarrow \infty$ where $M = (m\pm\Delta)^{2}$, the ignored potential term $\sim O(1/r)$ in Eq.\eqref{SysOde} will not generate more singular terms than $r^{2M}\exp(-Br^{2}/4)$ in power series expansion near $r \rightarrow \infty$. Note that the exciton vortex in the uniform orbital magnetic case carries the same charge and statistics as the case without the orbital field due to the same form of the effective topological field theory despite of the different forms for the zero mode solutions.

\section{Conclusion}
In summary, we have studied a thin film of topological insulator with both $T$-breaking Zeeman mass and $T$-symmetric excitonic mass. The two masses compete with each other and result in two topologically distinct phases for the elementary excitations: quantum anomalous Hall (QAH) and quantum spin Hall (QSH).  We studied the effective theory for the electromagnetism and exciton order parameter vortices by integrating out the fermions, and there can be other topological properties such as a helical metal at a particular kind of domain wall of the exciton order parameter. We also obtained explicit wave functions for the fermion zero mode at the vortices. There is one zero mode for QSH and no mode in the QAH regime, and found the zero mode solution under a uniform magnetic field in some limits. 

Before finishing this paper, we would like to emphasize some relevant facts for practical observation of exciton condensation in a topological insulator (TI)  thin film. In principle, TI might host a higher transition temperature $T^{*}$ for the exciton superfluid-insulator transition than graphene, resulting from the decreased number of fermion flavors.  In graphene, there has been intense theoretical study of the exciton condensates, and the estimated transition temperature for graphene ranges from milliKelvin~\cite{GrapheneSF2} to room temperature~\cite{GrapheneSF1}. An important reduction comes from the large number of fermion species in the screening~\cite{GrapheneSF2} ($N=8$ for bilayer graphene) which induces the factor $T^{*}\sim e^{-16}E_{f}$ where $E_{f}$ is the fermi energy of the system even though the bare interaction energy scale (without screening) is not small $r_{s} = e^{2}/\epsilon v_{f} \sim 1$ in the graphene. However, the thin-film topological insulator contains only two species $N=2$ because we have only one Dirac fermion per layer. But we have the disadvantageous situation for $r_{s}$ as in many current materials the dielectric $\epsilon$ for TIs are quite large (ranging from $30$ to $80$) though $v_{f}\sim 10^{5} m/s$ which is nearly half of graphene case.  (An additional problem of early TI materials, that they were not in practice very insulating, seems to have been overcome.)  These two factors contribute to reduced values $r_{s}\sim 1/3 - 1/8$. However, we can look for higher $r_{s}$ and thus higher transition temperature by noting that the main drawback comes from the large $\epsilon$ which is material-dependent; a possible solution is to sandwich a thin normal insulator of small $\epsilon^{*}$ between two topological insulators (See Fig.\ref{Fig1} B), and which will change $r_{s} \rightarrow e^{2}/\epsilon^{*} v_{f}$ which could be as large as $3$ if $\epsilon^{*}\sim 10$. This might open a way to achieve a dramatic increase in the superfluid transition temperature and realize the surprising physics of the topological exciton condensate.

\acknowledgments
G.Y.C thanks S. Ryu, P. Ghaemi, and R.S.K. Mong for helpful comments. Funding support for this work was provided by FENA (G.Y.C) and NSF DMR-0804413 (J.E.M.). The authors acknowledge M. Franz for pointing out a mistake in the earlier version of this work and bringing our attention to the paper~\cite{MF}. 

\section*{APPENDIX}

We show that if the exciton mass has a ``real'' domain wall (i.e., passes through zero with some fixed phase), there is a topological helical state.  Without loss of generality, 
let the exciton mass $\Delta(y) \rightarrow \Delta_{0}>0$ for $y \rightarrow -\infty$, and $\Delta \rightarrow -\Delta_{0}<0$ for $y \rightarrow \infty$ with $\Delta(y) \rightarrow 0$ as $y \rightarrow 0$.  Then we have a helical edge state localized at $y=0$. In terms of the four-spinor, the helical states are $\Psi_{+}(k) \sim \exp(ikx) \exp(-|\int_{0}^{y}\Delta(y) dy|) (1,1,1,-1)^{T}$ with $E(k)= k$ and $\Psi_{-}(k) \sim \exp(ikx) \exp(-|\int_{0}^{y}\Delta(y) dy|) (1,-1,1,1)^{T}$ with $E(k)=-k$ and two fermionic states $\Psi_{+}(k)$ and $\Psi_{-}(-k)$ are the Kramer pair. These 1D state is protected when $T$- symmetry at the domain wall is conserved as for spin Hall edge states. Thus, this helical state localized at the domain wall reflects the underlying topological states. This spin Hall type physics can be best understood if we look at the matrix form of the Hamiltonian $H = \Psi^{\dagger} {\bf H} \Psi$ with the proper unitary transformation from the original bases, 
\begin{equation}
{\bf H} = 
\left[ 
\begin{array}{cccc} 
\Delta & p_{+} & 0 & 0 \\
p_{-} & -\Delta & 0 & 0 \\ 
0 & 0 & -\Delta & p_{+} \\
0 & 0 &p_{-} & \Delta
\end{array} 
\right]
\end{equation}
As the helical 1D metal is protected only if $T$-symmetry is conserved at the domain wall, we set $m=0$ near the domain wall. Now, it's clear that when $\Delta$ has the domain wall at $y=0$, the upper two spinor has the chiral mode propagating in the positive $x$, and the lower two spinor has the chiral mode propagating in the negative $x$. Further consideration shows that two modes are a Kramer pair and are protected from generating a gap only when $T$-symmetry is conserved.   Note that a general domain wall where the phase difference in the (generally complex) exciton mass is not $\pi$ need not have this bound state.

\bibliography{topopart_final.bib}

%
%
%
%
%
%
%
%

\end{document}